## Title

Network analysis with quantum dynamics clarifies why photosystem II exploits both chlorophyll *a* and *b*

**Short title: The benefit of heterogeneity in a photosystem II**

## Authors

Eunchul Kim[1,2]†, Daekyung Lee[3]†, Souichi Sakamoto[4]†, Ju-Yeon Jo[4], Mauricio Vargas[5], Akihito Ishizaki[4,6]*, Jun Minagawa[1,2]*, Heetae Kim[3]*

## Affiliations

[1]Division of Environmental Photobiology, National Institute for Basic Biology; Myodaiji, 444-8585, Okazaki, Japan.
[2]Basic Biology Program, Graduate Institute for Advanced Studies, SOKENDAI, Okazaki 444-8585, Japan.
[3]Department of Energy Engineering, Korea Institute of Energy Technology, Naju 58330, Korea
[4]Institute for Molecular Science, National Institutes of Natural Sciences, Okazaki 444-8585, Japan
[5]Instituto de Matemáticas, Universidad de Talca, Casilla 747, Talca, Chile
[6]Molecular Science Program, Graduate Institute for Advanced Studies, SOKENDAI, Okazaki 444-8585, Japan

†These authors contributed equally to this work.
*To whom correspondence may be addressed. hkim@kentech.ac.kr (H.K.), minagawa@nibb.ac.jp (J.M.), ishizaki@ims.ac.jp (A.I.)

## Abstract

In green plants, chlorophyll-*a* and chlorophyll-*b* are the predominant pigments bound to light-harvesting proteins. While the individual characteristics of these chlorophylls are well understood, the advantages of their coexistence remain unclear. In this study, we establish a method to simulate excitation energy transfer within the entire photosystem II supercomplex by employing network analysis integrated with quantum dynamic calculations. We then investigate the effects of the coexistence of chlorophyll-*a* and chlorophyll-*b* by comparing various chlorophyll compositions. Our results reveal that the natural chlorophyll composition allows the excited energy to preferentially flow through specific domains that act as safety valves, preventing downstream overflow. Our findings suggest that the light-harvesting proteins in a photosystem II supercomplex achieve evolutionary advantages with the natural chlorophyll-*a*/*b* ratio, capturing light energy efficiently and safely across various light intensities. Using our framework, one can better understand how green plants harvest light energy and adapt to changing environmental conditions.

## Teaser

**Network analysis reveals that land plants can take functional advantage from heterogeneous chlorophylls in its photosystem.**

## Introduction

Chlorophylls (Chls) play a crucial role in photosynthesis by serving as the primary light harvesters that transfer energy to the reaction centers (RCs) of a photosystem, where charge separation takes place (*1*). Under conditions of excess light, however, Chls also participate in the thermal dissipation of excess excitation energy to prevent photooxidative damage (*2*). The light-harvesting system thus needs to be designed for both efficient light capture and effective photoprotection.

In land plants, a photosystem II supercomplex (PSII SC) contains peripheral light-harvesting antennae, each composed of three monomeric light-harvesting complex II (LHCII), CP24, CP26, and CP29, and two LHCII trimers, M-LHCII and S-LHCII (Fig. 1A) (*3*). In this photosystem, while both Chls *a* and *b* are in LHCIIs, only Chl *a* exists in the core complex. In fact, Chl *b* can increase the light-harvesting efficacy of LHCIIs because of its higher absorption coefficient at specific wavelengths and a higher excited energy level than that of Chl *a*. Therefore, it might be desirable for LHCIIs to exclusively bind Chl *b* to achieve maximal excitation energy transfer (EET) efficiency due to the increased energy gap between Chls in LHCIIs and PSII core complexes. However, natural LHCIIs of land plants still maintain both Chl *a* and *b*, the reason for which has yet to be elucidated.

Since the determination of the crystal structure of LHCII (*4*), the EET dynamics among the pigments in this structure have been studied using various experimental and theoretical approaches. By combining quantum chemical and electrostatic methods, researchers have calculated excitonic couplings and site energies of Chls, showing good agreement with experimental data (*5*, *6*). These results indicate a strong coupling between Chl *b* and Chl *a*, facilitating rapid excitation energy transfer from Chl *b* to Chl *a*. Furthermore, the existence of long-lived quantum coherence in an LHCII (*7*) and other photosynthetic antenna proteins such as the Fenna-Matthews-Olson (FMO) complex (*8*), has been revealed through two-dimensional electronic spectroscopy. This long-lived quantum coherence is proposed to enhance the energy transfer efficiency to the reaction center at physiological temperature (*9*). Recent studies have identified the importance of a noise-canceling network model for EET dynamics (*10*). Molecular mechanisms of photoprotective excitation quenching in LHCII have also been proposed, including the interaction between Chl and lutein (*11*) and the formation of charge transfer states within the Chl-carotenoid (Car) (*12*, *13*) and Chl-Chl (*14*, *15*) pairs. However, these analyses have focused on individual antenna protein molecules, with limited extension to the entire PSII SC using simplified interactions (*16*, *17*). This limitation arises from the complex nature of the photosynthetic antenna system, which involves numerous pigments and prevents the application of conventional computational methods typically used for smaller proteins such as the combination of quantum mechanics (QM)/molecular mechanics (MM) with molecular dynamics (MD) simulations (*18*) or coarse-grained approaches (*17*, *19*).

While the maturation of Cryo-EM technology, near-atomic structures of many photosystem SCs, ranging from land plants to various algae, are now accessible (*20*). Addressing EET dynamics across the entire SC considering the structural information of these SCs could tackle previously unreachable critical issues. For example, analyzing how the mixed composition of Chls within LHCII affects its function as PSII antennae, as studied in this report, is one such issue. Since both Chl *a* and *b* are crucial for forming natural LHCII (*21*), it is challenging to experimentally produce LHCII with various Chl compositions, making a computational approach more feasible.

This study presents a multidisciplinary approach to analyze the holistic EET dynamics across the entire PSII SC (*22*). Multiple EET domains were introduced by considering the quantum mechanical interactions within the LHCII molecules and PSII complexes. The holistic EET dynamics were then investigated by network science (*23*) that has been rapidly developed for the last 20 years to understand complex interactions (*23*, *24*), including neuroscience (*25*), power engineering (*26*), sociology (*27*), and ecology (*28*). By modeling the EET as a network, we were able to disclose and visualize how energy flows between different Chl domains, which is essential to understand the efficiency and regulation of the light-harvesting system of the PSII SC. This approach represents one of the most feasible methods currently available to investigate the holistic EET dynamics among numerous Chls in the entire PSII SC, clarifying the contributions of individual components to the EET dynamics. Finally, based on various versions of EET networks that vary the ratio of Chls *a* to *b*, the advantages of the natural Chl composition in the PSII SC are discussed.

## Results and Discussion

### The EET network of the natural PSII SC

To generate the natural EET network, we first estimated the EET rate constants between Chls in the natural PSII SC based on quantum dynamical methods. In this study, we analyzed the PSII SC (pdb. 5xnl) of *Pisum sativum*, for which coordinates of all Chls a and b belonging to the SC are known with detailed structural information (*19*). The EET rate depends on the site energies of Chls and the delocalized exciton states between strongly coupled Chls (Fig. 1B). To appropriately simulate the EET between Chls for a wide range of distances, we adopted the concept of 'domain' that corresponds to a group of strongly coupled Chls (*29*, *30*). For proximal EET within a domain, we applied Redfield theory, which has been widely employed for more accurately describing EET as the relaxation process of the excitons (*31*). The generalized Förster theory, which has been commonly employed to describe long range EET (*29, 32, 33*), was applied to consider further EET between delocalized exciton states belonging to different domains.

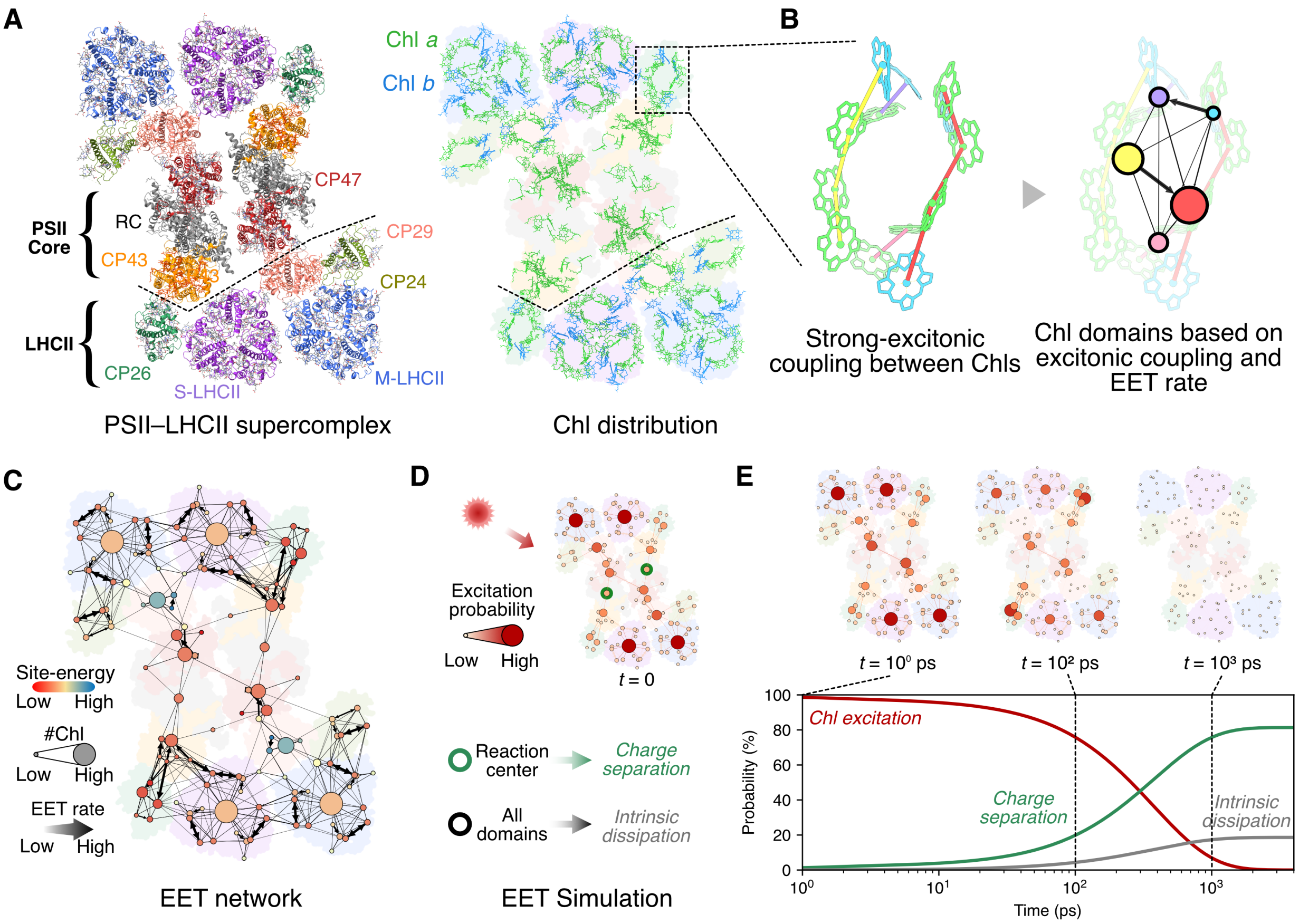

**Fig. 1. EET network analysis of natural PSII SC. (A)** Protein compositions and Chl distribution of the PSII SC. **(B)** Schematic representations of the Chl domains generated via excitonic coupling between Chls. **(C)** EET network and site energies of Chl domains in the natural PSII SC. The size and color of the circles represent the number of Chls in the domain and its averaged site energy. The direction and line width of the arrows between the circles represent the direction and the relative rate constants for EET. The rate constants larger than 200 $ps^{-1}$ are shown. **(D)** Excitation probability of domains at t=0 due to random Chl excitation for EET simulation considering the charge separation by the special pairs and intrinsic dissipation processes of Chls. **(E)** Simulated ensemble probabilities of the excited states of Chls, charge separation, and intrinsic dissipation after initial excitation.

We treated the EET information derived from this combined approach as a "network" and characterized it through network theory (*20*), which enables quantifying and simulating the dynamic process as interactions of "nodes" along with their topological characteristics (see "Methods"). In our EET network, the nodes are Chl domains, and the links represent EET rate constants connecting domains. In Fig. 1C, the color of each node represents the average site energies of Chls within each domain, revealing that the domains in CP29 and CP26 have the maximum (blue) and minimum (red) site energies, respectively. The EET links show that CP43 is tightly coupled to CP26 and S-LHCII, whereas CP47 is weakly coupled to peripheral LHCIIs. In this manner, the EET network can show how the PSII core is energetically integrated with each LHCII component of the peripheral antennae.

Excitation probability of each domain and its charge separation yield was estimated as an EET network, incorporating experimentally determined rate constants for the charge separation and intrinsic rate constants for the dissipation of electronically excited Chls through nonradiative and radiative pathways (Fig. 1D). We numerically simulated the dynamic behavior of EET as a Markov process, wherein the EET process is only influenced by the current states of the Chls or their domains (and not by past events) (*34*). By assuming that Chls in all domains are randomly excited at time zero, the excitation probability of each domain, charge separation, and intrinsic dissipation were calculated, which revealed the detailed dynamics of energy conversion from absorbed light energy to charge separation in a PSII SC (Fig. 1E, Supplementary Fig. S1). The distribution map of the excitation probability shows localization of excitation energy in CP26 at approximately 100 ps and that the decay lifetime of Chl excitation was estimated as 500 ps, comparable to experimentally determined values (490 ps for intact spinach PSII SC at pH 5.5) (*35*). The charge separation yield was determined to be 0.81, which is consistent with the range (0.78–0.84) of the maximum quantum yield experimentally determined in C3 plants (*36*). This consistency of the quantum yield between experiments and our simulation supports our approach as effective in determining the holistic excitation energy transfer dynamics within the entire PSII SC.

### Synthetic EET networks of hypothetical PSII SC models

To understand the advantages of the mixed Chl system of the natural PSII SC, we examined hypothetical PSII SC models consisting of various Chl *a* and Chl *b* combinations in LHCIIs (Fig. 2 and Supplementary Fig. S2). These models include all-*a* or all-*b* type LHCII that solely binds Chl *a* or Chl *b*, respectively. The position and orientation of Chls are the same as those of natural Chls. Furthermore, Chl *a* has a larger magnitude of transition dipole moment than Chl *b* (Supplementary Table S1). Due to stronger excitonic coupling between Chls *a* than Chls *b*, a domain in an all-*a* type LHCII has more Chls and has lower site energy than a domain in all-*b* type LHCII (Fig. 2A and C and Supplementary Table S2–11).

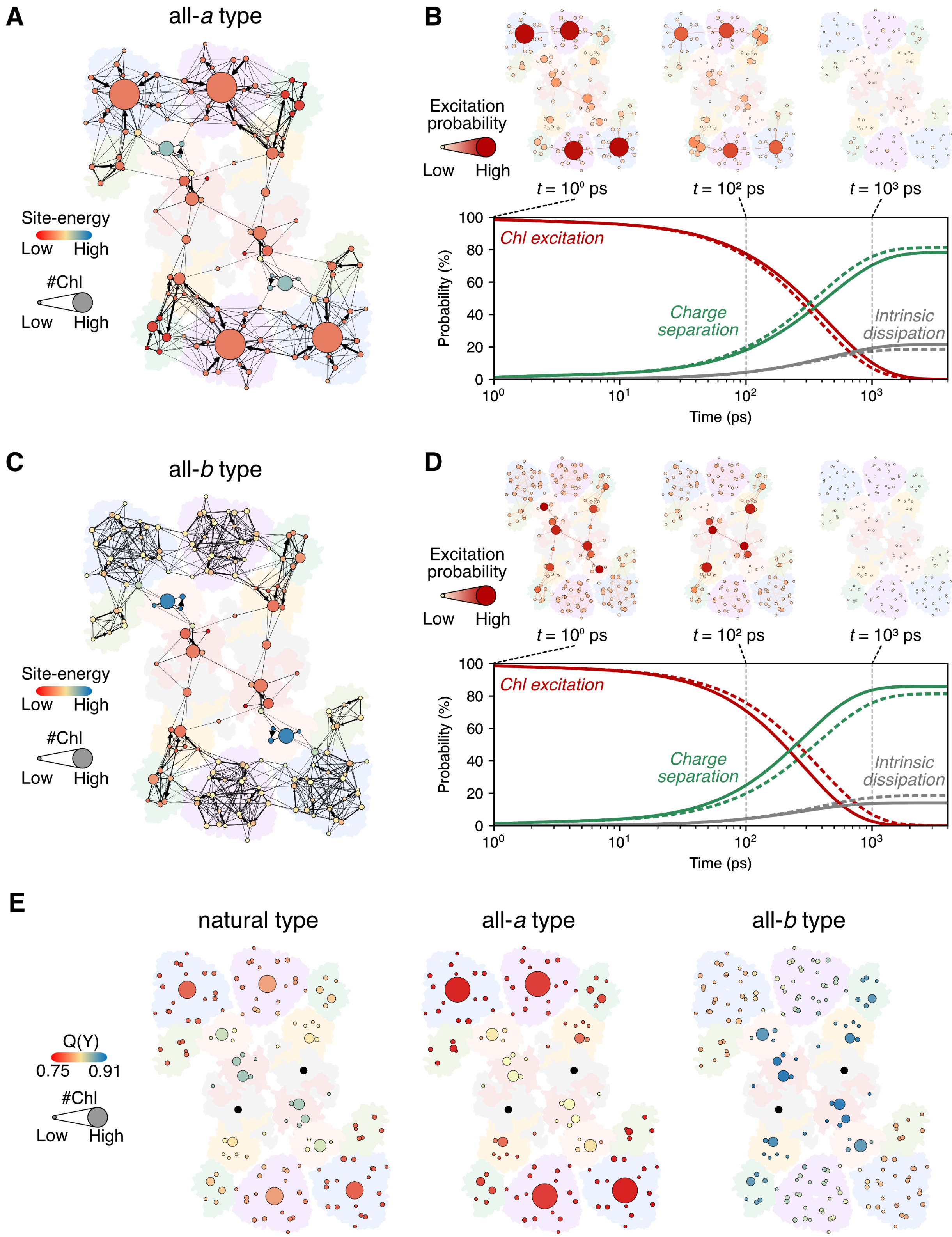

**Fig. 2. EET network analysis of all-*a* and all-*b* type PSII SCs. (A, C)** The EET network and site energies of Chl domains in the all-a and all-b type PSII SCs, respectively. The size and color of the circle represent the number of Chls in the domain and its averaged site energy. The direction and line width of the arrows between circles represent the direction and relative rate constants for EET. The rate constants larger than 200 $ps^{-1}$ are shown. **(B, D)** Simulated ensemble probabilities of the excited states of Chls, charge separation, and intrinsic dissipation after initial

excitation of a Chl in the all-*a* and all-*b* type PSII SCs, respectively. Solid lines represent the simulation results of either all-*a* or all-*b* type PSII SCs, and dashed lines represent the results of the natural PSII SC. **(E)** Individual yield, Q(Y), of domains in the natural, all-*a* and all-*b* type PSII SCs. Q(Y) of the domain represents the yield of charge separation when the domain is excited. Q(Y) of RC domains was not included in the scale map of Q(Y).

From the network analysis on the EET between the domains of the synthetic models, we also found that the decay of the excitation probabilities is slower in the PSII SC with all-*a* type LHCII (all-*a* type PSII SC) and faster in the PSII SC with all-*b* type LHCII (all-*b* type PSII SC) than in the natural PSII SC (Fig. 2B and D). Specifically, in the all-*a* type PSII SC, the excitation probabilities in the peripheral antennae, including S-LHCII, M-LHCII, and CP26, remained higher than those in the other types even at 1 ns (Fig. 2B and Supplementary Fig. S3). In all-*b* type PSII SC, however, the excitation probabilities in the inner antennae, including CP43 and CP47, were higher than those in the other area (Fig. 2D, Supplementary Fig. S4), leading to the highest charge separation yield among the comparative PSII SC models. These results indicate that the yield of charge separation at RCs is increased in all-*b* type PSII SC by the effect of the increased number of Chl *b* on the energy gap between LHCIIs and the PSII core, which suppresses the energy backflow from the PSII core to LHCIIs.

To further characterize the integrity of antennae, we estimated the probability of charge separation at the RC when each domain is excited individually, which we call individual yield, Q(Y) (Fig. 2E). Each colored circle corresponds to a domain with its color representing Q(Y). We found that Q(Y)s were higher in the PSII core than in the peripheral LHCIIs for all types of PSII SCs. This is due to the structure of the PSII SC in which the peripheral LHCIIs surround the PSII core. It makes the EET directed toward the RC, generating the gradient of Q(Y) from the peripheral LHCIIs to the PSII core. Interestingly, compared to the natural type PSII SC, the domains in the PSII core in the all-*a* type PSII SC featured lower Q(Y)s, and those in the all-*b* type PSII SC showed higher Q(Y)s. Note that all types of PSII SCs have an identical PSII core having only Chl *a*. Therefore, we emphasize that this occurs due to the large energy gap between LHCIIs and the PSII core, which more obstructs the energy backflow from the PSII core to LHCIIs in the all-*b* type PSII SC than the all-*a* type. In other words, 'the funneling effect' of EET is the most prominent effect in the all-*b* type PSII SC, which aligns well with previously reported characteristics of EET in the Fenna–Matthews–Olson complex such as the advantage of 'a ratchet system' (*37*). Therefore, after excitation energy is captured by the PSII SC, it can be efficiently utilized for charge separation at RCs.

**Cumulative excitation energy flow analysis of EET**

To trace excitation energy flow in PSII SCs, we propose two key metrics: cumulative input (CI) for each node and net cumulative flow (NCF) on each link. The CI represents the total accumulated energy input to node $j$, $\sum_{i} f_{ij}$,during the simulation, where $f_{ij}$ is the accumulated energy flow from node $i$ to $j$. Concurrently, the NCF between nodes $i$ and $j$ is defined by $|f_{ij} - f_{ji}|$, illustrating the net energy transfer dynamics between them. Based on the CI and the NCF values, we can reveal the energy transfer pathways in the PSII SC. For example, a node with a high CI value will serve as a gateway, where excitation energy frequently passes through, and the link with a high NCF value will be the major path of the energy traversal. The energy flow map in Fig. 3A shows the characteristic energy pathway of the natural type PSII (thick and dark arrows in Fig. 3A). The RCs mainly receive excitation energy from the two distinct in-flows through the domains in CP43 or CP47. In the peripheral area, where S-LHCII, M-LHCII, CP24, and CP26 reside, certain domains (red and large nodes) facilitate the building of particular pathways that contribute to frequent EET. These EET paths occur as a consequence of the site-energy landscape in the PSII SC related to the spatial position of low site-energy Chls. In the all-*a* type PSII SC, for instance, only a few domains feature high CI, and the EET pathways (distinguished by the dark and thick arrows) are denser in the peripheral region than in the natural region (Fig. 3B). Therefore, the energy flow to RCs can be interrupted and trapped in the peripheral area, resulting in a lower charge separation yield than that of the natural type. In the all-*b* type PSII SC, conversely, high-CI domains are positioned along the major EET pathways so that the energy flow can be effectively guided to RCs, which enables the highest yield of charge separation (Fig. 3C). Note that the high-CI domains have low site energy in PSII SC.

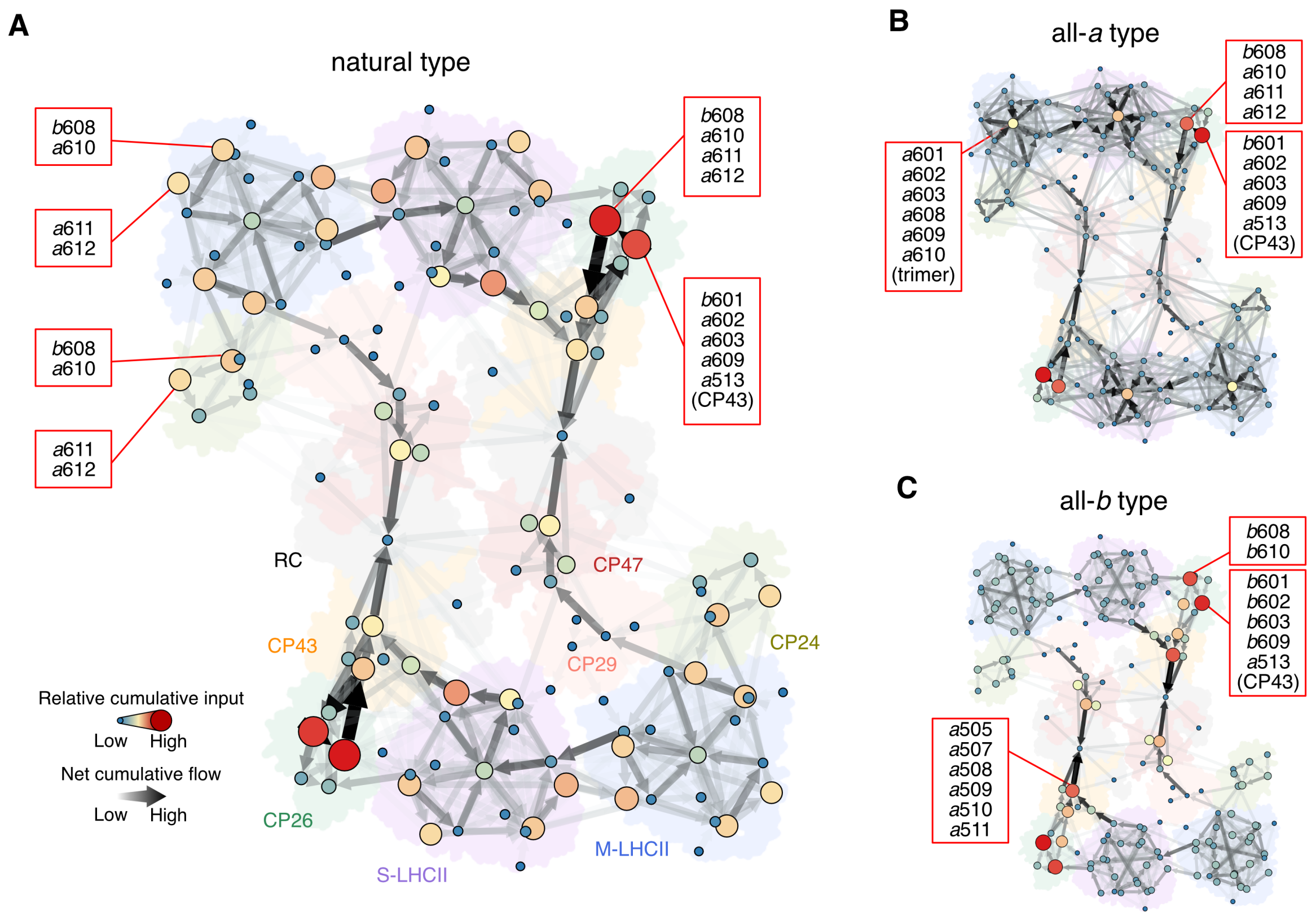

**Fig. 3. Cumulative flow analysis of PSII SCs. (A-C)** Relative cumulative input (CI) of each domain and net cumulative flow of each link in the natural, all-*a* and all-*b* PSII SCs, respectively. The Chls in the high CI domains are shown in the red boxes for each domain.

## Light-harvesting and photoprotective capabilities of PSII SCs

The increased amount of Chl *b* can improve the charge separation yield of the PSII SC beyond the increased absorption in several spectral areas of Chl *b*. Considering the solar radiation spectrum that both Chls *a* and *b* can absorb, having more Chl *b* in an LHCII can naturally achieve higher light-harvesting efficiency (Supplementary Fig. S5). However, we found in the PSII SC models that the yield is even higher (Fig. 4A, gray dots) than the expected natural yield advantage (Fig. 4A, dashed gray line). This indicates that the funneling effect described above enables the PSII SC to outperform the inherent advantage of having more Chl *b* in the antennae.

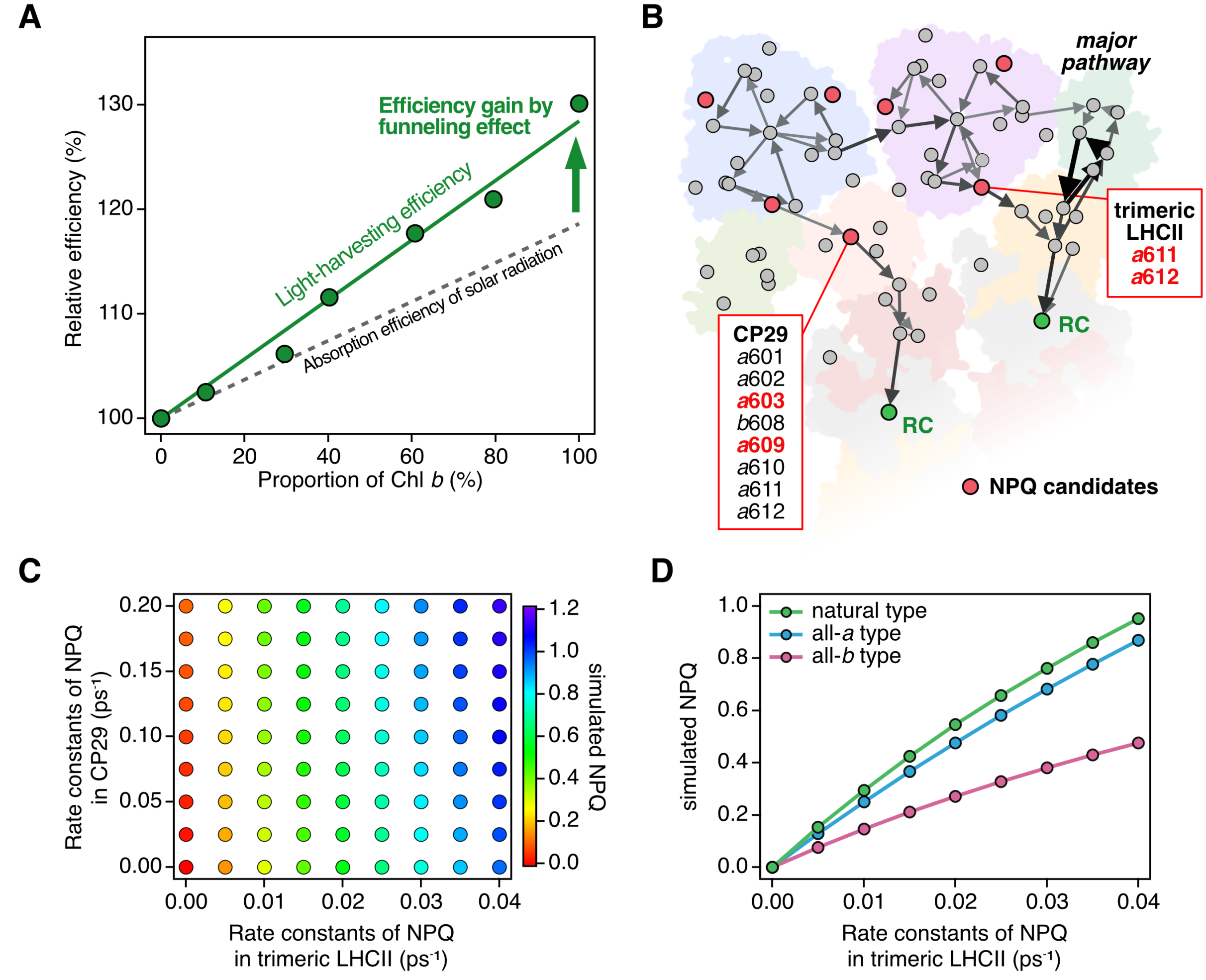

**Figure 4. Light-harvesting and photoprotective capabilities of PSII SCs. (A)** The relationship between the proportion of Chl b in the antenna of the PSII SCs compared with the relative absorption of solar radiation (dashed gray line) and the consequent relative net light-harvesting efficiency (green circles and line). **(B)** NPQ (nonphotochemical quenching) candidate domains (red circles) and RCs (green circles) and the major pathway of energy flow in the natural type PSII SC. Major pathways selectively represent the top 1% net cumulative flow. **(C)** Simulated NPQ values ([yield w/o NPQ] / [yield w/ NPQ] – 1) of the natural type PSII SC with various combinations of the rate constants of NPQ in trimeric LHCII and CP29 (Table S12). **(D)** Simulated NPQ values of the natural, all-*a* and all-*b* type PSII SCs with the rate constants of NPQ in trimeric LHCII (Table S12–14).

In general, the higher the yield of a PSII SC is, the larger the amount of light energy a plant can harvest. However, it is always beneficial only for plants under moderate light conditions. In nature, plants are often exposed to excessive light conditions that could cause photodamage to the photosystems (*38*). Thus, any photoprotection schemes are crucial for plants to avoid such photodamage for their survival under changing light conditions.

Our results also support that the natural PSII SC has great potential to prevent photodamage by non-photochemical quenching (NPQ) in trimeric LHCIIs or CP29s. In the natural PSII SC, we found that excitation energy mostly flows through specific domains along the major pathways in the LHCIIs (Fig. 4B). Intriguingly, these stepping domains include Chls *a*611-*a*612 and *a*603-*a*609, which have been proposed as photoprotective quenching sites in LHCIIs. Chls *a*611-*a*612, observed in S-LHCII and M-LHCII, have been proposed to quench the excitation energy by transferring the energy to lutein (*11*, *39*) or by charge transfer quenching (*14*). Chls *a*603-*a*609, observed in CP29, have been proposed to conduct a similar role by charge transfer quenching with zeaxanthin (*13*). Therefore, these stepping domains in the natural PSII SC can dissipate excessive energy, when necessary, as if firefighters were near fire. To address this hypothesis, we estimated the yield of PSII SC when NPQ process is activated and calculated the NPQ values by applying the various rate constants (0.005~0.04 $ps^{-1}$ for trimeric LHCII (*11*, *14*, *15*, 40, *41*) and 0.005~0.2 $ps^{-1}$ for CP29 (*13*, *42–44*)) of NPQ in trimeric LHCIIs or CP29s (Fig. 4C and D, Supplementary Table S12–14). The result showed that NPQ in trimeric LHCII much more efficiently quenches the excitation energy than that in CP29 (Fig. 4C). In the comparison of the PSII SCs, the simulated NPQ in natural and all-*a* type PSII SCs showed about twice higher values than that of all-*b* type PSII SC (Fig. 4D). This result supports that the natural type PSII SC forms an efficient structure to dissipate the excitation energy while maintaining high light-harvesting efficiency.

**Enhancing light-harvesting efficiency through the ratchet system**

Our results also provide the insight that the 'ratchet' system is advantageous over a monotonous funnel system for light harvesting efficiency. To consider the ratchet effect on EET, we investigated other models of PSII SC with various ratios of Chl b to Chls: trimerLHCII-*a*, trimerLHCII-*b*, monomerLHCII-*a*, and monomerLHCII-*b*. The trimerLHCII-*a* or monomerLHCII-*a* types involve the hypothetical trimeric or monomeric LHCIIs that bind only Chl *a*, respectively. Likewise, the trimerLHCII-*b* and monomerLHCII-*b* types are defined in the same way, binding only Chl *b*. Notably, the trimerLHCII-*a* type showed higher quantum yields than the monomerLHCII-*a* type, as well as the monomerLHCII-*b* type to the trimerLHCII-*b* type (Fig. 4A). We hypothesize that these features originated from the different excited energy levels of the Chls in trimeric LHCIIs and monomeric LHCIIs. In the trimerLHCII-*a* type and the monomerLHCII-*b* type, the overall excited energy levels of Chls in the monomeric LHCIIs are higher than those of the trimeric LHCIIs, which can cause the ratchet effect because the excitation energy in the core complexes can rarely migrate to the trimeric LHCIIs in this condition, and vice versa in the monomerLHCII-*a* type and the trimerLHCII-*b* type (Supplementary Fig. S2). This finding is in line with a previous study showing that the Fenna–Matthews–Olson (FMO) complex forms the ratchet system rather than the monotonous funnel system (*37*). This insight supports that the utilization of the ratchet system can be a potential strategy to increase light-harvesting efficiency in light-energy conversion systems such as photovoltaic materials, including solar cells.

Our EET analysis of the PSII SC stands out from previous studies by offering a holistic approach to understanding the entire system of PSII SC. By incorporating network analysis, we provide a comprehensive view of the EET process, focusing on the interplay between antenna complexes and core complexes within the EET network. Our analysis successfully simulated the 'light-harvesting' and 'photoprotection' capabilities of the PSII SC, highlighting the advantages of the coexistence of Chls *a* and *b* in natural LHCII, particularly when compared to hypothetical PSII SCs. Only the natural PSII SC efficiently transfers excitation energy to the RC while simultaneously protecting itself.

The EET network analysis can be further extended to investigate various characteristics of EET in photosynthetic protein complexes. This approach represents a novel step toward understanding the photosystem from a network perspective, with focus primarily on the topological characteristics and EET dynamics of the PSII SC. Further research can explore other network science approaches to this problem, including consideration of community characteristics and resilience analysis, which hold great promise for further comprehending natural systems.

## Materials and Methods

### Domain Formation

In large photosynthetic antenna systems, all embedded pigments are not necessarily strongly coupled. This coupling results in the formation of discrete domains, each of which comprise delocalized exciton states of strongly coupled pigments. To define these domains quantitatively, the threshold value is introduced. Pigment pairs $(m, n)$ whose electronic coupling $V_{mn}$ is greater than the threshold assigned to the same domain. The Redfield theory is often employed to describe exciton relaxation within the domains, whereas the generalized Förster theory (*29*, *32*, *33*) is a suitable choice to describe energy transfer between different domains. The electronic excitation localized on the $m$-th pigment and the *M*-th delocalized exciton state in the *d*-domain are denoted by $|m_d\rangle$ and $|M_d\rangle$, respectively. The corresponding electronic transition energies are written as $\epsilon_{m_d} = \hbar\omega_{m_d}$ and $\epsilon_{M_d} = \hbar\omega_{M_d}$. For convenience, $\omega_{\alpha\beta} = \omega_\alpha - \omega_\beta$ is introduced for any possible paris $(\alpha, \beta)$. Delocalized exciton states can be expressed by a linear combination of localized electronic excitations as $|M_d\rangle = \sum_{m_d} c_{m_d}^{(M_d)} |m_d\rangle$. It is noted that the subscription will be omitted when domains are not necessarily specified. The threshold value is usually chosen to be in the order of the environmental reorganization energy $E_\lambda$ to consider environment-induced dynamic localization effects. A typical value for the reorganization energy associated with electronic excitations in pigment-protein complexes is in the range (*37*) of $10\ cm^{-1} < E_\lambda < 100\ cm^{-1}$, and therefore, we assume the threshold value to be 30 $cm^{-1}$. We note that both Redfield and generalized Förster theories are computationally less expensive but quantitatively less accurate than more sophisticated theories (*31*). However, both the theories provide qualitatively reasonable results in the parameter region corresponding to natural photosynthetic systems, as was demonstrated in Ishizaki and Fleming (2009) (*31*). Therefore, we think it is relevant to employ the Redfield theory and the generalized Förster theory to investigate energy transfer dynamics in such a large system as the PSII SC. For the same reason, we will employ the point-dipole approximation to evaluate the excitonic couplings of pigment instead of more sophisticated theories (*45*, *46*).

The values of electronic excitation energies and excitonic couplings of pigments in the PSII core, LHCII, CP26, and CP29 are taken from Refs. (*6*, *47–49*). Because the parameters for CP24 have not been reported, they are substituted by those for the LHCII monomer in a similar fashion as in Ref. (*17*). The interprotein excitonic couplings are calculated with the use of the point-dipole approximation with the effective magnitudes of transition dipole moments (*50*),

$$V_{mn} = \frac{1}{4\pi\epsilon_0}\left[\frac{\vec{\mu}_m \cdot \vec{\mu}_n}{R_{mn}^3} - 3\,\frac{(\vec{\mu}_m \cdot \vec{R}_{mn})(\vec{\mu}_n \cdot \vec{R}_{mn})}{R_{mn}^5}\right], \tag{1}$$

where $\epsilon_0$ is the vacuum permittivity and $\vec{R}_{mn}$ is the center-to-center spatial vector between pigments *m* and *n*. The vector $\vec{\mu}_m$ stands for the transition dipole moment of the $Q_y$ band in chlorophyll. The direction is assumed to be along the $N_B$-$N_D$ axis (*29*), and the effective magnitudes are employed to include the influence of the surrounding protein environment (*50*). The X-ray structural data, 5XNL.pdb, are employed to evaluate the orientations of the transition dipole moments and distance vectors. The employed values are presented in Supplementary Table S1.

To describe the influence of the protein environment upon the involved electronic excitations, we consider the spectral density. Significant experimental and computational efforts have been made to obtain spectral density functions for various photosynthetic pigment-protein complexes, which provides the information on protein-induced fluctuations and intramolecular vibrational modes affecting the electronic states of embedded pigments. For computationally reproducing absorption and emission spectra of molecules, both the effects are necessary. As was demonstrated in Fujihashi et al. (2015) (*51*), however, the impact of the intramolecular vibrational modes on the energy transfer dynamics is typically eradicated by the protein-induced fluctuations at physiological temperatures. For this reason, we use the spectral density with no vibrational sideband (*52*),

$$J(\omega) = \frac{S}{s_1 + s_2} \sum_{i=1,2} \frac{s_i}{7!\, 2\omega_i^4} \omega^3 exp\left[-\left(\frac{\omega}{\omega_i}\right)^{1/2}\right] \tag{2}$$

with $S = 0.65$, $s_1 = 0.8$, $s_2 = 0.5$, $\omega_1 = 0.5565\ cm^{-1}$ and $\omega_2 = 1.936\ cm^{-1}$. The reorganization energy associated with electronic excitations is thus obtained with the well-known formula, $E_\lambda = \int_0^\infty d\omega \hbar\omega\, J(\omega)$, yielding the values of $E_\lambda = 50.87\ cm^{-1}$.

**Intradomain exciton transfer**

Among the pigments in a domain, we assume that the relaxation of one delocalized exciton state $|M\rangle$ to another $|N\rangle$ is described with the Redfield theory. Hence, the rate constant of exciton transfer is obtained as Ref. (*29*)

$$k_{M\to N} = 2\pi\gamma_{MN}\omega_{MN}^2\{[1 + n_{BE}(\omega_{MN})]J(\omega_{MN}) + n_{BE}(-\omega_{MN})J(-\omega_{MN})\}, \tag{3}$$

where $n_{BE}(\omega)$ is the Bose–Einstein distribution function. The factor $\gamma_{MN}$ is computed with the exciton coefficients $\{c_m^{(M)}\}$ as

$$\gamma_{MN} = \sum_{mn} e^{-R_{mn}/R_c} c_m^{(M)} c_m^{(N)} c_n^{(N)} c_n^{(M)}, \tag{4}$$

where $R_{mn}$ is the center-to-center distance between pigments $m$ and $n$, and $R_c$ is the correlation radius of the protein-induced fluctuations. We set $R_c = 5.0$ Å for numerical calculations (*52*). The inverse lifetime of state $|M\rangle$ is thus obtained as $\tau_M^{-1} = \frac{1}{2}\sum_N k_{M\to N}$, yielding the lifetime broadening in optical spectra. See Eqs. (7) and (8).

**Interdomain exciton transfer**

Energy transfer between delocalized exciton states each of which belongs to different domains is described as Förster-type transfer. The rate constant of the transfer from $|M_a\rangle$ to $|N_b\rangle$ $(a \neq b)$ is thus computed as

$$k_{M_a\to N_b} = 2\pi \frac{V_{M_aN_b}^2}{\hbar^2} \int_{-\infty}^{\infty} d\omega\, F_{M_a}(\omega) A_{N_b}(\omega). \tag{5}$$

In the equation, $V_{M_aN_b}$ represents excitonic coupling between delocalized exciton states,

$$V_{M_aN_b} = \sum_{m_a n_b} c_{m_a}^{(M_a)} c_{n_b}^{(N_b)} V_{m_a n_b}, \tag{6}$$

and $F_{M_a}(\omega)$ and $A_{N_b}(\omega)$ are the fluorescence and absorbance lineshape functions,

$$F_{M_a}(\omega) = \frac{1}{2\pi}\int_{-\infty}^{\infty} dt\, e^{-i(\omega-\tilde{\omega}_{M_a})t} e^{G_{M_a}(t)-G_{M_a}(0)} e^{-|t|/\tau_{M_a}}, \quad (7)$$

$$A_{N_b}(\omega) = \frac{1}{2\pi}\int_{-\infty}^{\infty} dt\, e^{i(\omega-\tilde{\omega}_{N_b})t} e^{G_{N_b}(t)-G_{N_b}(0)} e^{-|t|/\tau_{N_b}}, \quad (8)$$

respectively. In the lineshape functions, the function $G_{M_d}(t)$ is given by

$$G_{M_d}(t) = \gamma_{M_d M_d} \int_0^{\infty} d\omega \left\{[1+n(\omega)]J(\omega)e^{-i\omega t} + n(\omega)J(\omega)e^{+i\omega t}\right\}, \quad (9)$$

and $\tilde{\omega}_{M_d}$ is the frequency renormalized with the diagonal and off-diagonal contributions of the exciton-vibrational coupling,

$$\tilde{\omega}_{M_d} = \omega_{M_d} - \frac{\gamma_{M_d M_d} E_\lambda}{\hbar} + \sum_{K_d \neq M_d} \gamma_{M_d K_d} \wp \int_{-\infty}^{\infty} d\omega \frac{\omega^2\{[1+n(\omega)]J(\omega) + n(-\omega)J(-\omega)\}}{\omega_{M_d K_d} - \omega}. \quad (10)$$

We assume that intradomain exciton relaxation processes are sufficiently fast in comparison to interdomain exciton transfer, thus, the rate of interdomain exciton transfer from domain *a* to *b* is calculated as in Ref. (*30*)

$$k_{a\to b} = \sum_{M_a N_b} p(M_a) k_{M_a \to N_b}, \quad (11)$$

where $p(M_a) = e^{-\epsilon_{M_a}/k_B T} / \sum_{M_a} e^{-\epsilon_{M_a}/k_B T}$ is the thermal distribution of an exciton state $|M_a\rangle$ when only domain *a* is considered. The parameter $k_B$ is the Boltzmann constant. An average over the static disorder in site energies needs to be taken to calculate the interdomain transfer rate constant in Eq. (11) (*47*). We assume that the independent variation of the site energies obeys a Gaussian distribution with full width at half maximum, $\Delta_{inh}$ for all pigments, and we perform Monte Carlo sampling. The employed values for $\Delta_{inh}$ are given in Supplementary Tables S15 and S16. The simulated absorption spectrum was compared with experimentally obtained data (Supplementary Fig. S6). The results showed that the peak positions and widths of the simulated spectrum were in good agreement with the experiment data, demonstrating the validity of the theoretical model and parameters employed. The discrepancy in the shorter wavelength region below 640 nm is due to the exclusion of the $Q_x$ and Soret bands and intermolecular vibrational sidebands from the simulations.

**Linear absorption spectrum**

The linear absorption spectrum is obtained as a sum of the spectrum in each domain as Ref. (*30*)

$$A(\omega) = \sum_d A_d(\omega), \quad (12)$$

$$A_d(\omega) \propto \omega \langle \Sigma_{M_d} |\vec{\mu}_{M_d}|^2 A_{M_d}(\omega) \rangle, \quad (13)$$

where $\vec{\mu}_{M_d}$ is the transition dipole moment of the exciton state $|M_d\rangle$ expressed as $\vec{\mu}_{M_d} = \Sigma_{m_d} c_{m_d}^{(M_d)} \vec{\mu}_{m_d}$ and the bracket $\langle \ldots \rangle$ stands for the average over the static disorder in the site energies.

**Site energies and excitonic couplings in hypothetical PSII SC models**

If the chlorophyll in the light-harvesting systems is replaced with another type of chlorophyll, the structural and physical properties of the chlorophyll, such as the positions, orientations, site energies, and excitonic couplings, are expected to change. When the EET dynamics in PSII SC models with varied ratios of Chl $b$ to Chl $a$ are investigated, the values of the site energies and excitonic couplings should be essentially evaluated based on quantum chemical calculations. However, the aim of this study was to explore the general characteristics of the EET dynamics affected by chlorophyll substitution rather than the details. Then, we assume that the positions and orientations of the chlorophylls are fixed, and we determine the values of the site energies and excitonic couplings in these hypothetical PSII SC models with the approximations described below.

The site energy of chlorophyll is determined based on its $Q_y$ transition energy under the influence of environmental degrees of freedom. In the PSII SC, because each chlorophyll is affected by these different environments, the shift of the $Q_y$ transition energy from the vacuum is different in each chlorophyll, even if the types of chlorophylls are the same. Thus, when the chlorophyll in the PSII SC is substituted into the other type, the magnitude of the change in the site energy is expected to be different at each site. However, we assume that this magnitude, $\Delta E$, is the same throughout the PSII SC and is estimated as the difference in the $Q_y$ transition energy between Chl $a$ and Chl $b$ in solvent. We therefore choose the excitation energies in diethyl ether obtained as 662 nm for Chl $a$ and 644 nm for Chl $b$, respectively, which leads to $\Delta E = 422$ $cm^{-1}$. In the all-$a$, monomerLHCII-$a$, and trimerLHCII-$a$ types, we subtract $\Delta E$ from the site energies of Chl $b$, whereas we add $\Delta E$ to the site energies of Chl $a$ in the all-$b$, monomerLHCII-$b$, and trimerLHCII-$b$ types.

When the distance between the chlorophylls is small compared to the spatial extent of their wave functions, the point-dipole approximation can be employed to compute the values of the excitonic couplings between them. Within the point-dipole approximation, if the transition dipole moment of the m-th chlorophyll in the natural PSII SC $\vec{\mu}_m$ is replaced with $\vec{\mu}_m{}'$, the modified excitonic coupling between the m-th and n-th chlorophylls $V'_{mn}$ is expressed as

$$V'_{mn} = \frac{|\vec{\mu}_m{}'|\,|\vec{\mu}_n{}'|}{|\vec{\mu}_m|\,|\vec{\mu}_n|} V_{mn}. \qquad (14)$$

We assume that Eq. (14) is valid for all pairs of the chlorophylls in the hypothetical models of the PSII SC. Generally, this approximation is not satisfied, and the evaluation of the excitonic couplings requires a more accurate method of quantum chemical calculation (*53*, *54*). However, for the excitonic couplings involving only the $Q_y$ transitions of chlorophylls in LHCII, it has been reported that the point-dipole approximation is sufficiently accurate to compute the values of these couplings (*55*). We assume that this fact is applicable to monomeric light-harvesting complexes, and we employ Eq. (14) to obtain the values of the excitonic couplings in the hypothetical models.

### Excitation energy transfer simulation

The consecutive excitation energy transfer is initiated at one of the domains and terminated when the excitation energy is utilized for charge separation at special pairs in RCs (photosynthesis) or is intrinsically dissipated. We denote the set of transient states of the 134 total domains as $T=\{i|i\leq 134\}$ and two absorbing states representing the charge separation and dissipation as $A=\{i|i=135, 136\}$. Representing the possible states S at each transition, the number of states $|S|=|T|+|A|=134+2=136$. In the photosystem, for all time $t \geq 0$ and all possible states $i_0, \ldots, i, j$, the excited energy transition process is a stochastic process that follows $\mathbb{P}(x_{t+1}=j|x_t=i, x_{t-1}=i_{t-1}, \ldots, x_0=i_0) = \mathbb{P}(x_{t+1}=j|x_t=i) = p_{ij}$ such that it satisfies the Markov property (*56*, *57*). Therefore, we

generate the transition rate matrix $Q \in \mathbb{R}^{|S| \times |S|}$ (also known as generator matrix) from the excited energy transition as a Markov process (*34*). Assuming $x(t_0) = i \in T$, for the first-order reaction of the excited energy transfer during photosynthesis, where the transition jump continues to the next state $j \in T$ at $t_1 \sim \exp(\lambda_{ij})$, we approximate the transition rate from the EET rate such that $q_{ij} \approx \lambda_{ij}$. Similarly, the rate constant of charge separation in RCs is set as $q_{ij} = 1/1.5\ \text{ps}^{-1}$(time constant = 1.5 ps) (*58*, *59*), where $i$ is the particular domain in RCs and $j$=135, while the dissipation is set identically for all domains $q_{ij} = 5\times10^{-4}\ \text{ps}^{-1}$ (time constant = 2 ns) (*60–62*), where $i \in T$ and $j$=136. In addition, the system can incorporate the process of NPQ. This process is characterized by the conversion of energy to thermal energy in specific domains, as illustrated in Fig. 4B, which identifies the domains belonging to trimeric LHCII and CP29. Therefore, in these domains, the usual dissipation rate for all nodes is supplemented with the additional dissipation rate due to NPQ. We simulated the system's yield with different NPQ rates assigned to the domains in LHCII and CP29, respectively, varying these rates to observe the consequences. The results are described in Fig. 4C and 4D.

A simulation was also performed with the rate constants (charge separation, 0.5~3 $\text{ps}^{-1}$; intrinsic dissipation, 1~4 $\text{ns}^{-1}$) and showed that the experimentally obtained values resulted in a consistent yield of charge separation (0.81), which is close to the experimentally obtained yield (Supplementary Fig. S7). Note that the transient states ($i \in T$) represent the distinct domains in PSII that physically exist, whereas the two additional absorbing states ($i \in A = \{135, 136\}$) are conceptual. The diagonal elements $q_{ii} = -\sum_{j \neq i} q_{ij}$ considering that no self-transition occurs. We then convert the transition rate matrix of PSII to the embedded discrete-time Markov chain, $P = I - Q/l$, where I is the identity matrix and $l$ is the smallest element of the diagonal of Q, $l = \min(q_{ii})$ (*63*). The elements of P, $p_{ij}$, indicate the probability of transferring the excitation energy between states. We consider the excited energy transition between S states such that $\sum_{j=1}^{S} p_{ij} = 1$ for all $i$.

Upon giving the initial conditions of the states, we numerically track the transition process of the excited energy by multiplying P. Let the vector $x(t) = [x_1(t), x_2(t), \ldots, x_{136}(t)]$ be the state vector indicating the excitation probability of each state $i$ at Markov time (or step) $t$. Since the vector represents the probability distribution, at all Markov times $t$, it satisfies $\sum_{i=1}^{S} x_i(t) = 1$. We let a domain $i$ be initially excited by defining the $i$-th element $x_i(0)=1$ and $x_{j \neq i}(0)=0$ otherwise in the initial state vector x(0). The excited energy can move to the next state following the transition probability $\mathbb{P}(x(t+1)=j \mid x(t)=i) = p_{ij}$, such that the next states $x(1)=x(0)P$; $x(2)=x(1)P=x(0)P^2$; …; $x(t)=x(0)P^t$. In this sense, we recursively estimate the probability distribution at any $t$-th step. Note that $x(0)P^t$ depends on the initial distribution of x(0). We assume that only a domain is initially excited, and no additional multiple excitation occurs during the Markov process. To estimate the average probability distribution over the initial target domains, we take the ensemble average over $i$, such that $\underline{x}(t) = \sum_{i}^{T} p_i x_{(i)}(0) P^t$, where $p_i$ is the initial excitation probability of domain $i$. In this study, we assume that the excitation probability of domain $i$ is proportional to the number of Chls, which is denoted by $h_i$. Therefore, we estimate the average probability distribution for PSII as $\underline{x}(t) = \frac{1}{h_{tot}} x_T(0) P^t$, where $x_T(0) = [h_1, h_2, h_3, \ldots, h_{133}, h_{134}, 0, 0]$ with $x_i = h_i$ for $i \in T$ or 0 for $i \in A$, and $h_{tot} = \sum_{i=1}^{T} h_i$. We further normalize $\underline{x}(t)$ by the random probability, 1/134, to analyze the probability distribution relative to the random event. We call the normalized average excitation probability of domains the excitation tendency $\rho_i(t) = 134 \times \underline{x}(t)$. The excitation tendency per protein is the average of that of domains: $\rho_p(t) = \rho_{ip}(t)$, where $p$ is the protein index.

### Network analysis of various PSII network models

We convert the natural PSII SC into a network structure, which we refer to as the natural PSII network, with nodes representing the domains in PSII and links for the energy transfer rate between the pairs of them. The spatial coordinates of domains remain in the PSII network, and the directed link weight from node $i$ to $j$ is identical to $p_{ij}$ in P. Therefore, the PSII network is a directed and weighted fully connected spatial network for the total number $n=|T|=134$ of nodes and the total number $m=n(n-1)=17{,}822$ of links. The PSII network is in principle a fully connected network between all pairs of domains, since the EET rates between domains are all positive values. However, considering the fact that the EET rate decays rapidly as the distance between Chls increases, we filter out the links with negligible low weight remaining a backbone of the network. In this study we estimate the energy transfer dynamics based on the fully connected relationship while we visualize the network with only some part of all links for visual clarity.

In the PSII network, where the excited energy transfers between domains are probabilistically proportional to the EET rate, total cumulative flow (TCF) analysis can reveal the apparent pattern of energy delivery from each node to the reaction center. To formally define the cumulative energy flow, we start by specifying the directed edge from $i$ to $j$ as $f_{ij} = \sum_{t=1}^{t_{max}} x_i(t)p_{ij}$, where $t_{max}$ represents the simulation time, which in our study is set to $10^5$. The net cumulative flow (NTF) of link $(i, j)$ is then defined as the absolute difference between $f_{ij}$ and $f_{ji}$, and follows the direction of the larger energy flow. Furthermore, we define the cumulative input (CI) of node $i$ as $\sum_k^T f_{ki}$, which corresponds to the total incident energy of the node. These formal definitions enable the accurate characterization of excited energy transfer within the PSII network.

For the comparative analysis, we constructed various types of PSII network models in addition to the natural PSII network. By alternating Chl *b* to Chl *a* (Chl *a* to Chl *b*) in an LHCII of the natural PSII network, we generate an all-*a* type (all-*b* type) PSII network that has only a single type of Chl in LHCII. We also constructed the intermediate versions of PSII networks between those models by partially modifying chlorophylls. Specifically, we change the Chls only for those that are in monomeric or trimeric components in LHCII. By this, we prepared seven distinct PSII networks varying the composition of the Chl types (Supplementary Fig. S2).

**Acknowledgments**

**Funding:**

JSPS KAKENHI Grant 23K14216 (E.K.)
JSPS KAKENHI Grant 23H04960 (E.K. and J.M.)
JSPS KAKENHI Grant 21H01052 (S.S. and A.I.)
JSPS KAKENHI Grant 21H05040 (J.M.)
National Research Foundation of Korea (NRF) grant funded by the Korea government (MSIT) NRF-2022R1C1C1005856 (D.L. and H.K.)
Korea Institute of Energy Technology Evaluation and Planning (KETEP) No. 20224000000320 (D.L. and H.K.)
Ministry of Trade, Industry & Energy (MOTIE) of the Republic of Korea, and No. 20224000000100 (D.L. and H.K.)
Human Frontier Science Program Award No. RGY0076, MEXT Quantum Leap Flagship Program Grant No. JPMXS0120330644 (A.I.)

**Author contributions:** E.K., D.L., S.S., J.-Y.J., M.V., A.I., J.M. and H.K. designed the research; E.K., D.L., S.S., J.-Y.J., A.I. and H.K. performed the research; A.I., J.M. and H.K. supervised the research; E.K., D.L., S.S., A.I., J.M. and H.K. wrote the paper.

**Competing interests:** Authors declare that they have no competing interests.

**Data and materials availability:** All data needed to evaluate the conclusions in the paper are present in the paper and/or the Supplementary Materials. Any additional data can be obtained upon request from the corresponding author.

## Figures and Tables

**Fig. 1. EET network analysis of natural PSII SC. (A)** Protein compositions and Chl distribution of the PSII SC. **(B)** Schematic representations of the Chl domains generated via excitonic coupling between Chls. **(C)** EET network and site energies of Chl domains in the natural PSII SC. The size and color of the circles represent the number of Chls in the domain and its averaged site energy. The direction and line width of the arrows between the circles represent the direction and the relative rate constants for EET. The rate constants larger than 200 $ps^{-1}$ are shown. **(D)** Excitation probability of domains at t=0 due to random Chl excitation for EET simulation considering the charge separation by the special pairs and intrinsic dissipation processes of Chls. **(E)** Simulated ensemble probabilities of the excited states of Chls, charge separation, and intrinsic dissipation after initial excitation.

**Fig. 2. EET network analysis of all-*a* and all-*b* type PSII SCs. (A, C)** The EET network and site energies of Chl domains in the all-a and all-b type PSII SCs, respectively. The size and color of the circle represent the number of Chls in the domain and its averaged site energy. The direction and line width of the arrows between circles represent the direction and relative rate constants for EET. The rate constants larger than 200 $ps^{-1}$ are shown. **(B, D)** Simulated ensemble probabilities of the excited states of Chls, charge separation, and intrinsic dissipation after initial excitation of a Chl in the all-*a* and all-*b* type PSII SCs, respectively. Solid lines represent the simulation results of either all-*a* or all-*b* type PSII SCs, and dashed lines represent the results of the natural PSII SC. **(E)** Individual yield, Q(Y), of domains in the natural, all-*a* and all-*b* type PSII SCs. Q(Y) of the domain represents the yield of charge separation when the domain is excited. Q(Y) of RC domains was not included in the scale map of Q(Y).

**Fig. 3. Cumulative flow analysis of PSII SCs. (A-C)** Relative cumulative input (CI) of each domain and net cumulative flow of each link in the natural, all-*a* and all-*b* PSII SCs, respectively. The Chls in the high CI domains are shown in the red boxes for each domain.

**Figure 4. Light-harvesting and photoprotective capabilities of PSII SCs. (A)** The relationship between the proportion of Chl b in the antenna of the PSII SCs compared with the relative absorption of solar radiation (dashed gray line) and the consequent relative net light-harvesting efficiency (green circles and line). **(B)** NPQ (nonphotochemical quenching) candidate domains (red circles) and RCs (green circles) and the major pathway of energy flow in the natural type PSII SC. Major pathways selectively represent the top 1% net cumulative flow. **(C)** Simulated NPQ values ([yield w/o NPQ] / [yield w/ NPQ] – 1) of the natural type PSII SC with various combinations of the rate constants of NPQ in trimeric LHCII and CP29 (Table S12). **(D)** Simulated NPQ values of the natural, all-*a* and all-*b* type PSII SCs with the rate constants of NPQ in trimeric LHCII (Table S12–14).

## Supplementary Materials

Figs. S1 to S7
Tables S1 to S16
Reference (*64*)